\documentclass[aps,twocolumn]{revtex4}
\usepackage[dvips]{graphics,graphicx}
\usepackage{amsmath}

\begin{document}

\title{Josephson oscillation in the dissipative Bose-Hubbard dimer}
\author{ Andrey R. Kolovsky$^{1,2}$}
\affiliation{$^1$Kirensky Institute of Physics, 660036, Krasnoyarsk, Russia} 
\affiliation{$^2$Siberian Federal University, 660041, Krasnoyarsk, Russia} 
\date{\today}
\begin{abstract}
We analyze Josephson's oscillation of Bose particles in the open (dissipative) Bose-Hubbard dimer. First, we excite the dimer from the vacuum state into a state suitable for observing the oscillation by using a special protocol for external driving. Next, we switch off the driving and observe the oscillation. It is shown that the main mechanism for the decay of Josephson's oscillation is the dephasing due to fluctuating number of particles in open systems. An analytical estimate for the decay time is obtained.
\end{abstract}
\maketitle

The conservative two-site Bose-Hubbard model or the Bose-Hubbard (BH) dimer is, perhaps, the most studied system of identical particles. It serves as the model for the Josephson oscillation (periodic change in occupations of the lattice sites) and the self-trapping (interaction-induced destruction of the inter-site tunneling) \cite{Smer97,Kell02,Albi05,Gati07}, semiclassical quantization of the many-body systems \cite{Mahm05,Grae07}, highly correlated states like fragmented condensates and NOON state \cite{Cira98,Spek99,Muel06,Garc12,110}, and the excited state quantum phase transitions \cite{Capr08,Lee09,Tren16}. From the experimental viewpoint the BH dimer can be realized by using several platforms among which are cold atoms in a two-well optical potential \cite{Albi05,Gati07,Tren16}, two coupled optical micro-cavities with Kerr nonlinearity \cite{Hart14,Cao16}, and two capacitively coupled transmons \cite{Raft14,Fedo21}. Notice that the last two systems do not conserve the number of particles and, thus, they should be described in the framework of the open or dissipative Bose-Hubbard model \cite{Kord15,116}. The particle exchange with reservoirs enriches the dynamics of the BH dimer, leading to several new effects which are not present in the conservative BH dimer. These are the resonant transmission \cite{123,Fedo21} and the quantum manifestation of bifurcations in the classical driven-dissipative dimer \cite{Cao16,Cast17,Gira21}. The latter research direction also includes analysis of the true steady-state of the system, as well as the metastable states which are responsible for hysteresis in the quantum case \cite{Rodr17,Ming18}.

In this work we address a particular problem of Josephson's oscillation (JO) in the open BH dimer \cite{remark}. The questions to answer are (i) the difference between JO in the conservative and open BH dimers and (ii) a suitable setup where one can observe the dissipative JO. We consider the system of two coupled quantum nonlinear oscillators where the first oscillator is excited by a monochromatic force while the second oscillator is subject to decay. Then the governing equation for the system density matrix $\widehat{{\cal R}}$ has the form
\begin{equation}
\label{1}
\frac{\partial \widehat{{\cal R}}}{\partial t}=-i[\widehat{{\cal H}}, \widehat{{\cal R}}] -\frac{\gamma}{2}
\left(\hat{a}_2^{\dagger}\hat{a}_2\widehat{\cal R }-2\hat{a}_2\widehat{\cal R }\hat{a}_2^{\dagger}
+\widehat{\cal R }\hat{a}_2^{\dagger}\hat{a}_2 \right) \;,
\end{equation}
where $\gamma$ is the relaxation constant. (Through the parer we set the fundamental Planck constant to unity, i.e., the commutation relation for the creation and annihilation operators is $[\hat{a}_\ell,\hat{a}_m^\dagger]=\delta_{\ell,m}$.) Using the rotating wave approximation the Hamiltonian $\widehat{{\cal H}}$ in Eq.~(\ref{1}) reads 
\begin{eqnarray}
\nonumber
\widehat{{\cal H}}= \Delta \sum_{\ell=1}^{2}\hat{n}_{\ell}
-\frac{J}{2}\left( \hat{a}_{2}^{\dagger}\hat{a}_{1} +{\rm h.c.} \right) \\
\label{2}
+\frac{U}{2}\sum_{\ell=1}^{2}\hat{n}_{\ell} (\hat{n}_{\ell}-1) + \frac{\Omega}{2}(\hat{a}_1^\dagger + \hat{a}_1) \;,
\end{eqnarray}
where $\Omega$ is the Rabi frequency, $\Delta$ the detuning, $J$ the coupling constant, and $U$ the microscopic interaction constant. To be certain, in what follow we assume $U\ge 0$.

\begin{figure}
\includegraphics[width=8.0cm,clip]{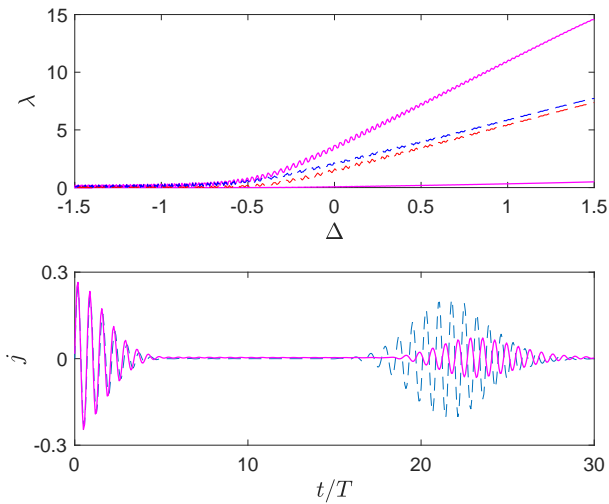}
\caption{Top panel: Eigenvalues of the matrix (\ref{rho}), solid lines, and its diagonal elements, dashed lines, as the functions of the detuning $\Delta$ during the adiabatic passage where $\Delta$ is changed linearly in time in the interval $|\Delta|\le 1.5$.  The system parameters are $J=0.5$, $U=0.25$, $\Omega=1/\sqrt{2}$, and $\gamma=0.002$. The sweeping rate is 50 tunneling periods $T=2\pi/J$ per unit interval of $\Delta$. Bottom panel: JO induced by switching of the driving at the end of the adiabatic passage. The dashed line is Eq.~(\ref{5}).} 
\label{fig1}
\end{figure}

Before considering JO we need to excite the system. We do this by capturing it into the nonlinear resonance \cite{1,Hann86,32}  (more precisely, into the limit cycle originated from the nonlinear resonance if $\gamma\ne0$) when we adiabatically sweep the detuning in the interval $\Delta_{in}< \Delta < \Delta_{f}$, where the initial $\Delta_{in}\ll -J$ and the final $\Delta_{f}$ is smaller than the critical $\Delta_{cr}\approx U(\Omega/\gamma)^2$ at which the basin of the limit cycle shrinks to zero \cite{preprint}.  The system dynamics during this adiabatic passage is illustrated in Fig.~\ref{fig1}(a).  Shown are the eigenvalues $\lambda_j$  and diagonal elements $\rho_{\ell,\ell}$ of the dimer single-particle density matrix $\hat{\rho}$,
\begin{equation}
\label{rho}
\rho_{\ell,m}(t)={\rm Tr}[\hat{a}^\dagger_\ell \hat{a}_m \widehat{{\cal R}}(t)] \;.
\end{equation}
It is seen in Fig.~\ref{fig1}(a) that after $\Delta\approx -J$ the number of bosons in the dimer is proportional to the detuning, $\rho_{\ell,\ell}\approx \Delta/U$. Also note that $\lambda_1\gg \lambda_2$. Thus, we have almost pure Bose-Einstein condensate with well-defined macroscopic wave function $\Psi(t)=[\psi_1(t),\psi_2(t)]$. For parameters of Fig.~\ref{fig1}(a) the amplitudes $\psi_\ell$ at the end of the adiabatic passage are $\psi_1=2.74$ and $\psi_2=2.66-0.013i$. Importantly, the amplitudes appear to be slightly different. This allows us to induce JO by simply switching off the driving. The bottom panel in Fig.~\ref{fig1} illustrates the evolution of the system for the next 30 tunneling periods after switching off the driving by showning the mean current $j(t)={\rm Tr}[\hat{j}\hat{\rho}(t)]$ where $\hat{j}$ is the single-particle current operator with the elements $j_{\ell,m} \sim (\delta_{\ell,m+1} -h.c.)/2i$. As expected we observe periodic oscillation of the current which is accompanied by oscillations of the occupation numbers (not shown) with some characteristic frequency $\omega$. Less expected is that this oscillation exhibits very rapid (as compared with the relaxation time $1/\gamma$) decay which is then followed by a revival. In the rest of the paper, we explain this effect and provide the estimate for the JO decay time $\tau$.  

As known, in the conservative BH dimer JO are described by the mean-field Hamiltonian  
\begin{equation}
\label{3}
H_{eff}=\frac{g}{2} I^2 - \frac{J}{2}\sqrt{1-I^2}\cos \phi \;,\quad g=UN \;,
\end{equation}
where $\phi$ is the phase difference between amplitudes $\psi_\ell$, $I$ the relative mismatch in the occupation numbers, $I=(|\psi_2|^2 - |\psi_1|^2)/(|\psi_2|^2 + |\psi_1|^2)$, and $g$ the macroscopic interaction constant. It follows from Eq.~(\ref{3}) that the frequency of small oscillation near the elliptic point $(I,\phi)=(0,0)$ is given by the relation
\begin{equation}
\label{4}
\omega = \sqrt{J^2+4Jg} \;.
\end{equation}
Equation (\ref{4}) correctly predicts the characteristic frequency $\omega$ if we define $g$ as $g=U\bar{N}$ where $\bar{N}$ is the mean number of bosons in the dimer (which, in the absence of driving, decays exponentially with the rate $\gamma$). However, Eq.~(\ref{4}) alone does not describe the rapid decay and revivals of JO. One finds the explanation for the latter effect by noticing that the excited state of the system is a state with fluctuating number of particles.  
\begin{figure}
\includegraphics[width=6.5cm,clip]{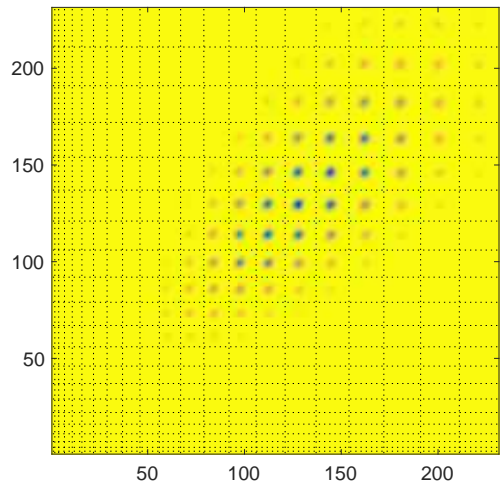}
\caption{ Total density matrix $\widehat{{\cal R}}$ at the end of the adiabatic passage. Shown are elements of the matrix by the absolute value. The dashed lines indicate dimensions of the Hilbert subspaces for the given $N$, where the truncation parameter is set to $N_{max}=20$.} 
\label{fig2}
\end{figure}

Figure \ref{fig2} shows the elements of the total density matrix $\widehat{{\cal R}}$ at the end of the preparation stage. The blocks along the main diagonal correspond to the `partial' density matrices $\widehat{{\cal R}}_N$ associated with the given number $N$ of bosons in the system. We denote traces of these matrices by $P_N(t)$. These quantities determine the probability to find $N$ particles in the dimer at a given time. Thus, we have the following equation for the mean current,
\begin{equation}
\label{5}
j(t)=\sum_{N=1}^\infty {\rm Tr}[\hat{j}_N\widehat{{\cal R}}_N(t)] \sim \sum_{N=1}^\infty P_N(t) \sin(\omega_N t) \;,
\end{equation}
where $\omega_N = \sqrt{J^2+4JUN}$. The current calculated by using Eq.~(\ref{5}) is depicted in Fig.~\ref{fig1}(b) by the dashed line. The nice qualitative agreement with the exact result confirms that the rapid decay of JO is due to the dephasing of oscillations for different $N$.  

Equation (\ref{5}) can be elaborated further by considering the limit of large $\bar{N}$ where $P_N(t)$ is approximated by the Gaussian, 
\begin{equation}
\label{6}
P_N \sim \exp\left[-\frac{(N-\bar{N})^2}{2\sqrt{\bar{N} }} \right] \;,\quad \bar{N}(t)=\bar{N}(0)\exp(-\gamma t) \;.
\end{equation}
The distribution (\ref{6}) follows from the estimate for the width $\delta N$ of the quantum nonlinear resonance, $\delta N\sim(\Omega\sqrt{\bar{N}} /U)^{1/2}$ \cite{1}. It should be opposed to the distribution $P_N \sim \exp[-(N-\bar{N})^2 /2\bar{N} ]$ for the coherent state, which would be the case if $U=0$. Thus, the state depicted in Fig.~\ref{fig2} is a squeezed state with the reduced fluctuations of the particle number. Using Eq.~(\ref{6}) we obtain  
\begin{equation}
\label{7}
j(t) \sim \exp\left(-\frac{t^2}{\tau^2}\right) \sin(\omega t) \;,\quad \tau^2 = \frac{4\omega^2}{ J^2 U^2 \sqrt{\bar{N}}} \;,
\end{equation}
where $\omega=\sqrt{J^2+4U\bar{N}}$.  It is also an appropriate place here to mention that the discussed effect is a formal analog of the decay and revival of the Rabi oscillation in the Jaynes-Cummings model \cite{Jayn63,Remp87}. However, the decay time of JO has different scaling law with the mean number of bosons in the system. 

To summarize, we extended the analysis of JO in the conservative BH dimer onto the dissipative BH dimer, where the mean number of particles in the system decreases exponentially with the rate $\gamma$. Of course, to observe JO  the relaxation rate $\gamma$ has to be much smaller than the Josephson frequency $\omega$. However, this condition is insufficient. Because of a fluctuating number of particles JO in the open BH dimer decay not as $\sim\exp(-\gamma t)$ but much faster, as $\sim\exp[-(t/\tau)^2]$, where $\tau$ is the dephasing time parametrized by the mean number of particles $\bar{N}$. For a moderate $\bar{N}$ one also observes revivals of JO for $t>\tau$. We also would like to stress the importance of the chosen initialization procedure since it establishes quantum correlations between the states with the different numbers of particles. These correlations are seen in Fig.~\ref{fig2} as the off-diagonal blocks. If the correlations were absent, it would be impossible to observe JO in the dissipative BH dimer in principle. 

The author thanks D. N. Maksimov for fruitful  discussions and acknowledges financial support from the Russian Science Foundation.



\begin{thebibliography}{99}
\bibitem{remark} 
The Josephson oscillation in BH dimer coupled to a reservoir was analyzed in Ref.~\cite{Sinh19}. However, the authors considered a model with no particle exchange. In this case, the effect of reservoir reduces to decoherence of the system density matrix (i.e., no dissipation).
\end{thebibliography}
\end{document}